\documentclass[sigconf]{acmart}
\renewcommand\footnotetextcopyrightpermission[1]{}

\AtBeginDocument{%
  }

\usepackage{tabularx,colortbl}
\usepackage{array}
\usepackage{booktabs}
\usepackage{multirow}
\usepackage{makecell}
\setcopyright{acmlicensed}

\begin{document}
\title{TriMoE: Augmenting GPU with AMX-Enabled CPU and DIMM-NDP for High-Throughput MoE Inference via Offloading}

\author{
    Yudong Pan$^{1,2}$, 
    Yintao He$^{1,2}$, 
    Tianhua Han$^{1,2}$, 
    Lian Liu$^{1,2}$, 
    Shixin Zhao$^{1,2}$, 
    Zhirong Chen$^{1,2}$, 
    Mengdi Wang$^{1,2}$, 
    Cangyuan Li$^{1,2}$, 
    Yinhe Han$^{1,2}$, 
    Ying Wang$^{1,2,\dagger}$
}

\affiliation{
  \institution{
    $^1$SKLP, Institute of Computing Technology, Chinese Academy of Sciences, Beijing, China \\
    $^2$University of Chinese Academy of Sciences, Beijing, China
  }
  \country{} 
}

\email{panyudong23@mails.ucas.ac.cn, heyintao19z@ict.ac.cn, wangying2009@ict.ac.cn}

\begin{abstract}

To deploy large Mixture-of-Experts (MoE) models cost-effectively, offloading-based single-GPU heterogeneous inference is crucial. While GPU–CPU architectures that offload cold experts are constrained by host memory bandwidth, emerging GPU-NDP architectures utilize DIMM-NDP to offload non-hot experts. However, non-hot experts are not a homogeneous memory-bound group: a significant subset of warm experts exists is severely penalized by high GPU I/O latency yet can saturate NDP compute throughput, exposing a critical compute gap. We present TriMoE, a novel GPU–CPU–NDP architecture that fills this gap by synergistically leveraging AMX-enabled CPU to precisely map hot, warm, and cold experts onto their optimal compute units. We further introduce a bottleneck-aware expert scheduling policy and a prediction-driven dynamic relayout/rebalancing scheme. Experiments demonstrate that TriMoE achieves up to 2.83$\times$ speedup over state-of-the-art solutions.

\end{abstract}

\maketitle
\pagestyle{plain} 
{\let\thefootnote\relax\footnotetext{$\dagger$ Corresponding author}}

\section{Introduction}

Mixture-of-Experts (MoE) models improve the efficiency–quality trade-off of large language models (LLMs) by sparsely activating a small set of experts per token~\cite{jiang2024mixtral, liu2024deepseek}. Yet the total parameters across hundreds of experts far exceed the memory of a single high-end GPU~\cite{eliseev2023fast, xu2025moe}. For example, serving DeepSeek-V2 requires at least seven 80GB H100 GPUs~\cite{fang2025klotski}, costing about \$200K. This makes offloading experts to host memory essential for single-GPU serving.

\begin{figure}
\centering
\includegraphics [width=1.0\linewidth]{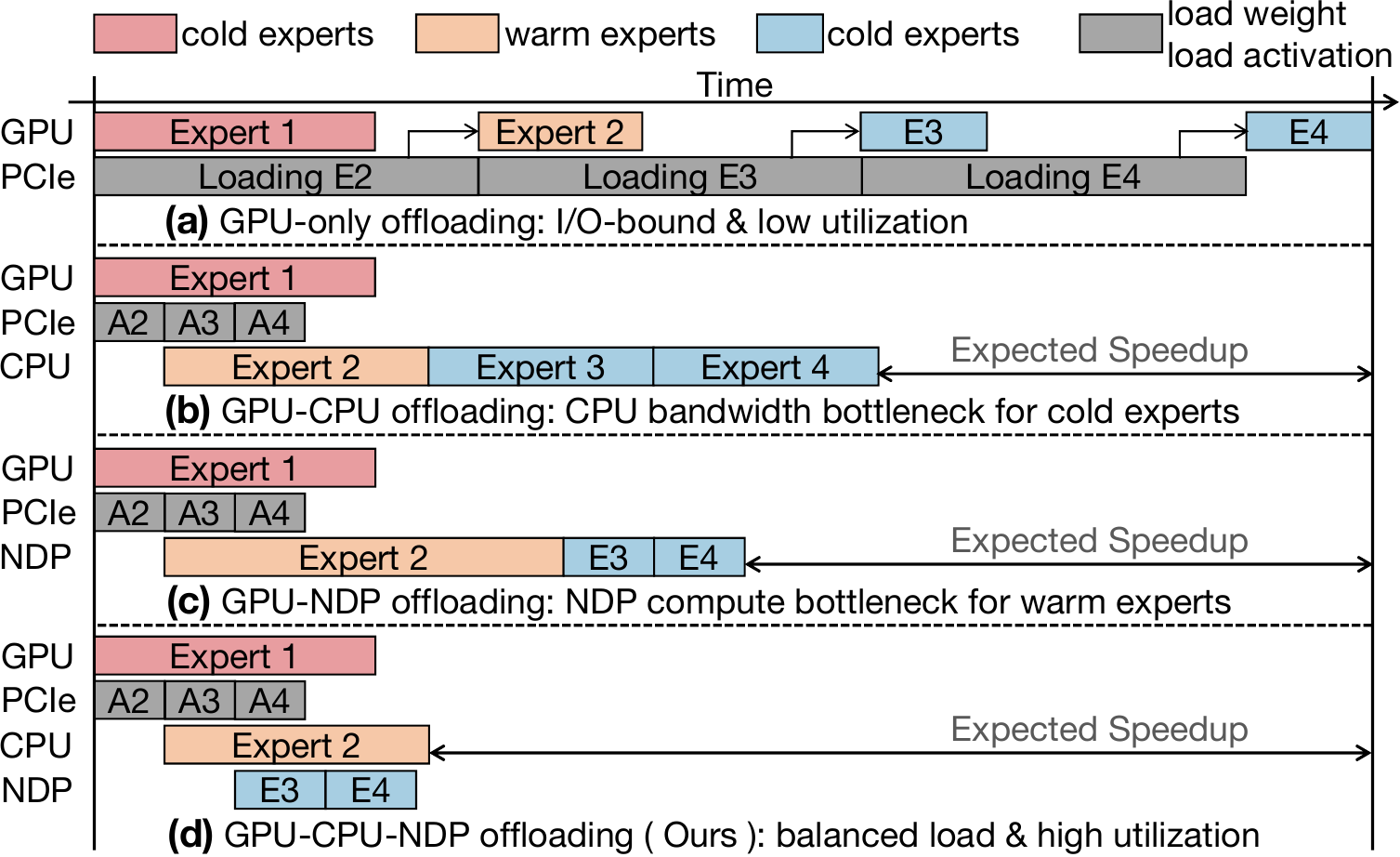}
\caption{Execution timelines of baseline MoE offloading systems and our proposed architecture.}
\label{fig:overall_compare}
\end{figure}

In MoE offloading systems, end-to-end latency is often dominated by data movement to fetch off-GPU expert weights over PCIe rather than pure computation. As illustrated in Figure~\ref{fig:overall_compare}, prior work therefore evolved from GPU-only prefetching to heterogeneous designs~\cite{kamahori2024fiddler}. Because GPU-CPU approaches are constrained by host memory bandwidth, current systems favor GPU-NDP~\cite{kim2024monde}, which exploits routing sparsity by keeping hot experts on the GPU and offloading low-load, memory-bound experts to NDP engines.



Revisiting expert activations under high-throughput workloads (e.g., offline~\cite{xu2025moe} or zigzag batching~\cite{fang2025klotski}), we find that the non-hot experts are not homogeneous. In addition to a small number of hot experts and a long tail of cold experts, ~30\% of experts are warm (a.k.a. mid-intensity). Mapping warm experts to the GPU yields insufficient tokens per expert for high utilization and fails to overlap PCIe transfers; mapping them to NDP exceeds the limited near-data compute budget, offsetting the gains from offloading cold experts. Therefore, the binary GPU-NDP partitioning paradigm, which lumps all non-hot experts into a single class, fails.



This analysis prompts a re-examination of host-side compute options. Although GPU-CPU systems have historically been viewed as constrained, we argue that this was due to the misallocation of CPU to bandwidth-intensive cold experts. In contrast, modern server CPUs with matrix extensions (e.g., Intel AMX and ARM SME) offer tens of TFLOPS~\cite{kim2025lia, chen2025ktransformers}, precisely matching the compute profile of warm experts. Meanwhile, DIMM-NDP remains the ideal choice for cold experts, leveraging high aggregate internal bandwidth that is 4–8$\times$ higher than that of the host memory system~\cite{liu2025l3, liu2025make}.

We therefore propose \textbf{TriMoE}, a single-GPU, high-throughput MoE inference architecture that uniquely orchestrates three heterogeneous compute domains within one system. TriMoE retains hot experts (e.g., shared experts) on the GPU to exploit peak throughput; maps warm experts to AMX-enabled CPU to bypass PCIe bottlenecks and match compute demand; and offloads long-tail cold experts to DIMM-NDP engines to leverage high internal bandwidth. This hierarchical design effectively shields the GPU from PCIe congestion and the CPU from memory-bound tasks, ensuring each domain focuses on the workload it handles most efficiently.

In a nutshell, our contributions are as follows:
\begin{itemize}
\item \textbf{Problem identification:} We are the first to identify and formalize the overlooked class of warm experts in high-throughput MoE serving and show why the frontier GPU-NDP offloading is intrinsically inefficient.
\item \textbf{TriMoE architecture \& scheduler:} We design TriMoE, the first offloading architecture to integrate GPU, AMX-enabled CPU, and DIMM-NDP, along with its scheduler that precisely matches experts to their most suitable compute units, thereby architecturally resolving fundamental scheduling conflicts.

\item \textbf{Dynamic relayout and rebalancing:} We design a prediction-driven policy that resolves the conflicting data layout preferences (GPU/CPU vs. NDP) via dynamic expert relayout, and prevents NDP load skew via cold expert rebalancing.
\item \textbf{End-to-end gains:} Across multiple representative MoE models, TriMoE achieves 2.12–2.83$\times$ speedups in decode, yielding 2.09–2.78$\times$ end-to-end throughput improvement over state-of-the-art offloading systems.
\end{itemize}
\section{Background}

\subsection{MoE Architecture \& Inference}
\begin{figure}
\centering
\includegraphics [width=0.83\linewidth]{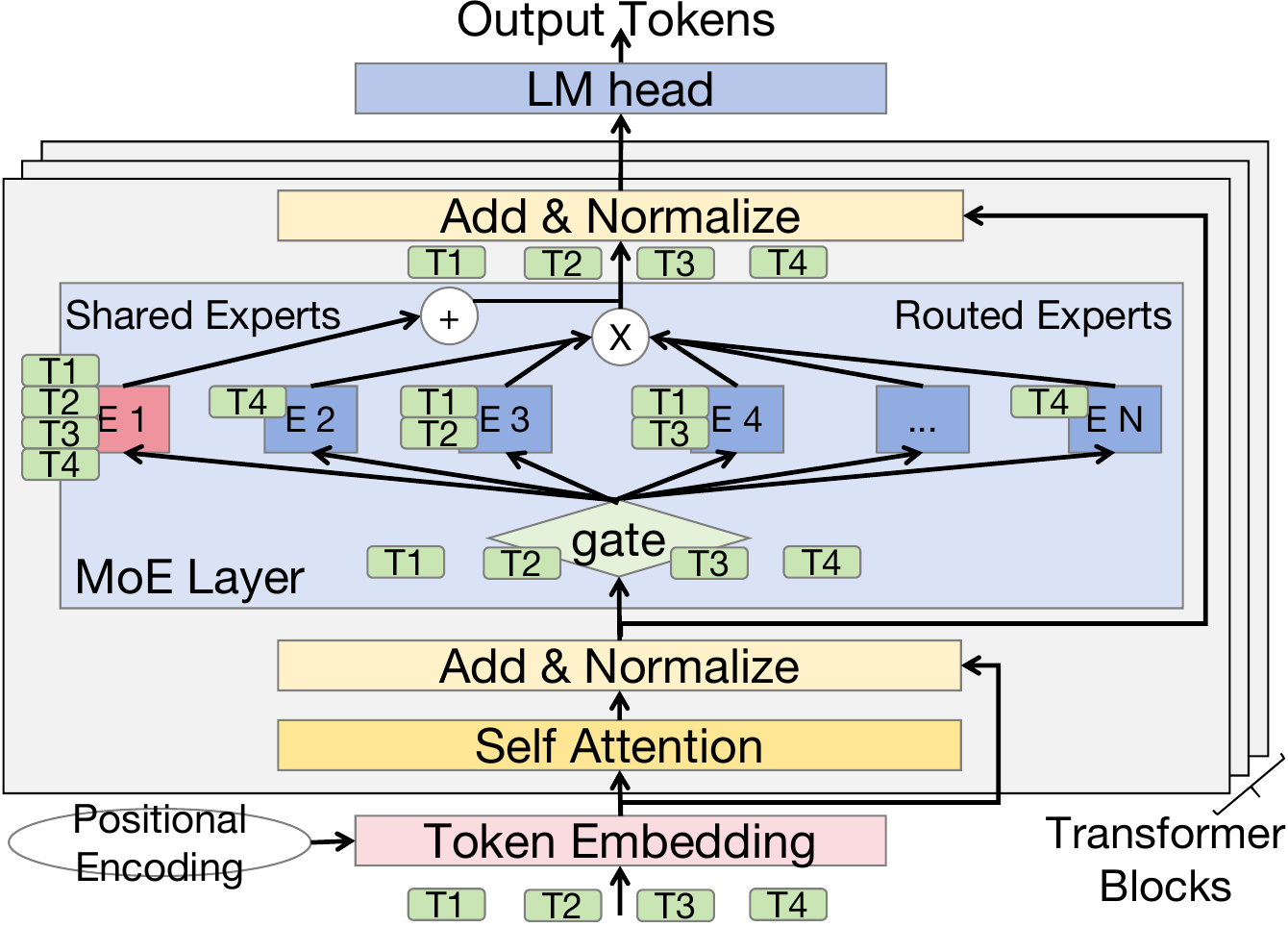}
\caption{MoE architecture with shared and routed experts.}
\label{fig:moe-structure}
\end{figure}

Mixture-of-Experts (MoE) models scale their parameter count without a linear increase in computational cost by sparsely activating expert subsets~\cite{jiang2024mixtral, liu2024deepseek, liu2024deepseek3, yang2025qwen3}. As shown in Figure~\ref{fig:moe-structure}, an MoE layer replaces the standard feed-forward layer with a gating function and N parallel experts. During inference, the gating function computes the assignment probability for each input token to every expert and selects top-k experts for processing. The outputs of these k experts are subsequently combined through a weighted sum to form the token's final output, which is then passed to subsequent layers. In practice, the total number of experts (N) often ranges from tens to hundreds, while the number of activated experts (K) is typically a small integer. Furthermore, modern architectures such as DeepSeek~\cite{liu2024deepseek} introduce both shared and routed experts. For instance, DeepSeek V2 utilizes 2 shared experts activated by all tokens, alongside 160 routed experts from which 6 are selected via a top-k gate. This hybrid design enables the model to achieve a good balance between parameter scale and computational cost.

\subsection{Offloading in MoE Inference}\label{Sec:MoE-offloading}
The large parameter scale of MoE models, due to their sparse architecture, typically necessitates costly multi-GPU deployments. Offloading-based inference systems have emerged as a cost-effective single-GPU solution~\cite{sheng2023flexgen, xue2024moe}, utilizing host memory to store most (>90\%) of expert weights. However, the PCIe bus (64 GB/s for PCIe 5.0) creates a severe I/O bottleneck, since its bandwidth lags GPU compute. Consequently, SOTA offloading systems target high-throughput scenarios (offline or zigzag batching) to amortize I/O overhead~\cite{xu2025moe, fang2025klotski}. To minimize expert data movement during the decode phase with highly skewed expert activation, designs have evolved from prefetching to heterogeneous computing paradigms. Hot experts are kept resident on the GPU, while other non-hot experts are executed on the host using its available compute resources~\cite{kamahori2024fiddler, zhong2025hybrimoe}. As heterogeneous GPU-CPU systems are still constrained by the host memory bandwidth, the growing trend is to utilize DIMM-NDP~\cite{kim2024monde, liu2025make}. For example, rank-level PUs leverage an aggregate internal bandwidth 4-8$\times$ higher than the host to process experts near-data~\cite{liu2025l3}, alleviating the I/O bottleneck and improving throughput. During the prefill phase, conversely, experts are widely activated, and GPU computation largely hides I/O overhead.

\begin{figure}
\centering
\includegraphics [width=1.0\linewidth]{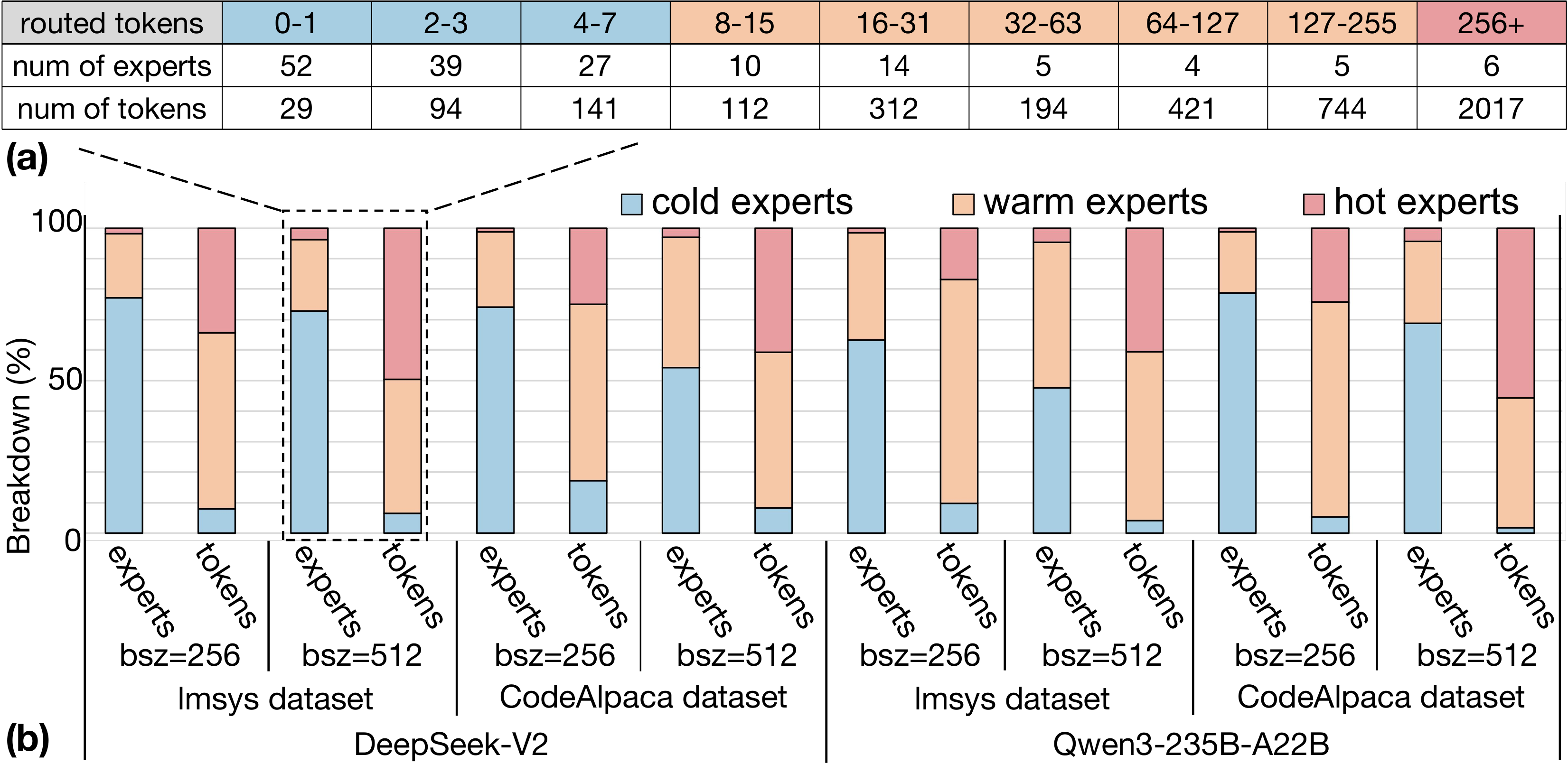}
\caption{Expert Activation across Batch Sizes, Models and Datasets\protect\footnotemark. (a) Fine-grained activation for a specific configuration. (b) Summary of activation across all configurations.}
\label{fig:hot-warm-cold}
\end{figure}
\footnotetext{``Batch size'' here refers to offline large-batch inference or zigzag batching that aggregates multiple small batches.}
\begin{figure*}
\centering
\includegraphics [width=0.8\linewidth]{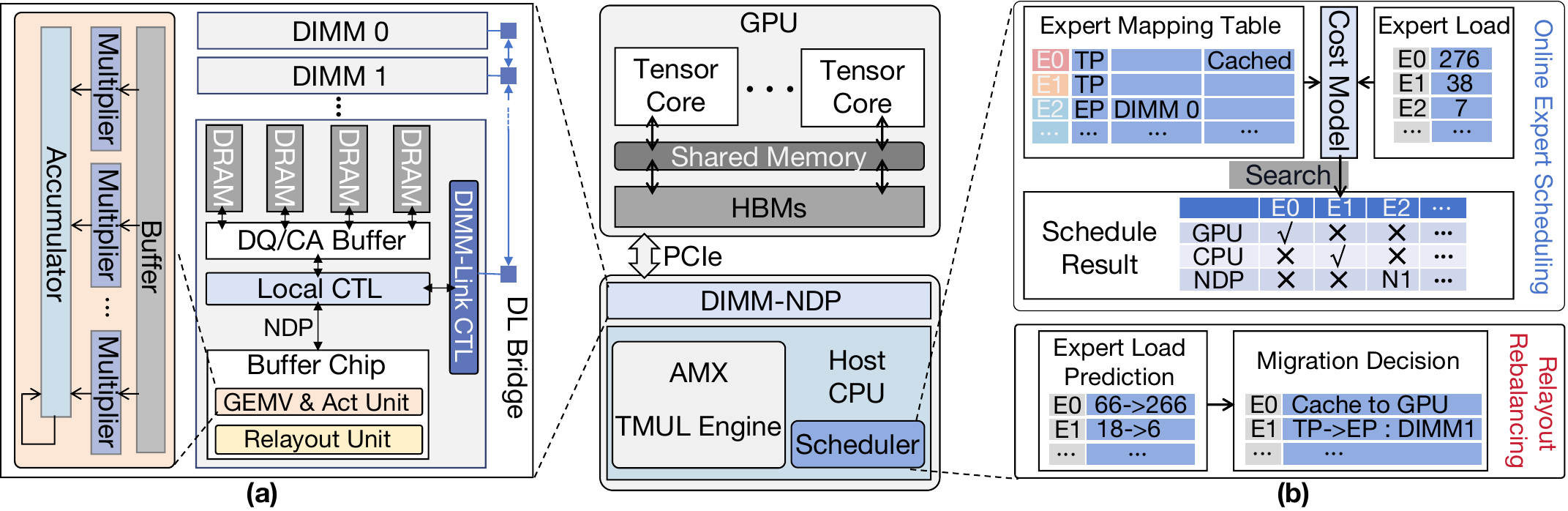}
\caption{Overview of TriMoE. (a) DIMM-NDP architecture. (b) Online expert scheduling \& Offline expert relayout-rebalancing.}
\label{fig:system}
\end{figure*}

\section{Motivation}

\subsection{The Scheduling Dilemma of Warm Experts}\label{Sec:Dilemma-of-warm}
\begin{figure}
\centering
\includegraphics [width=1.0\linewidth]{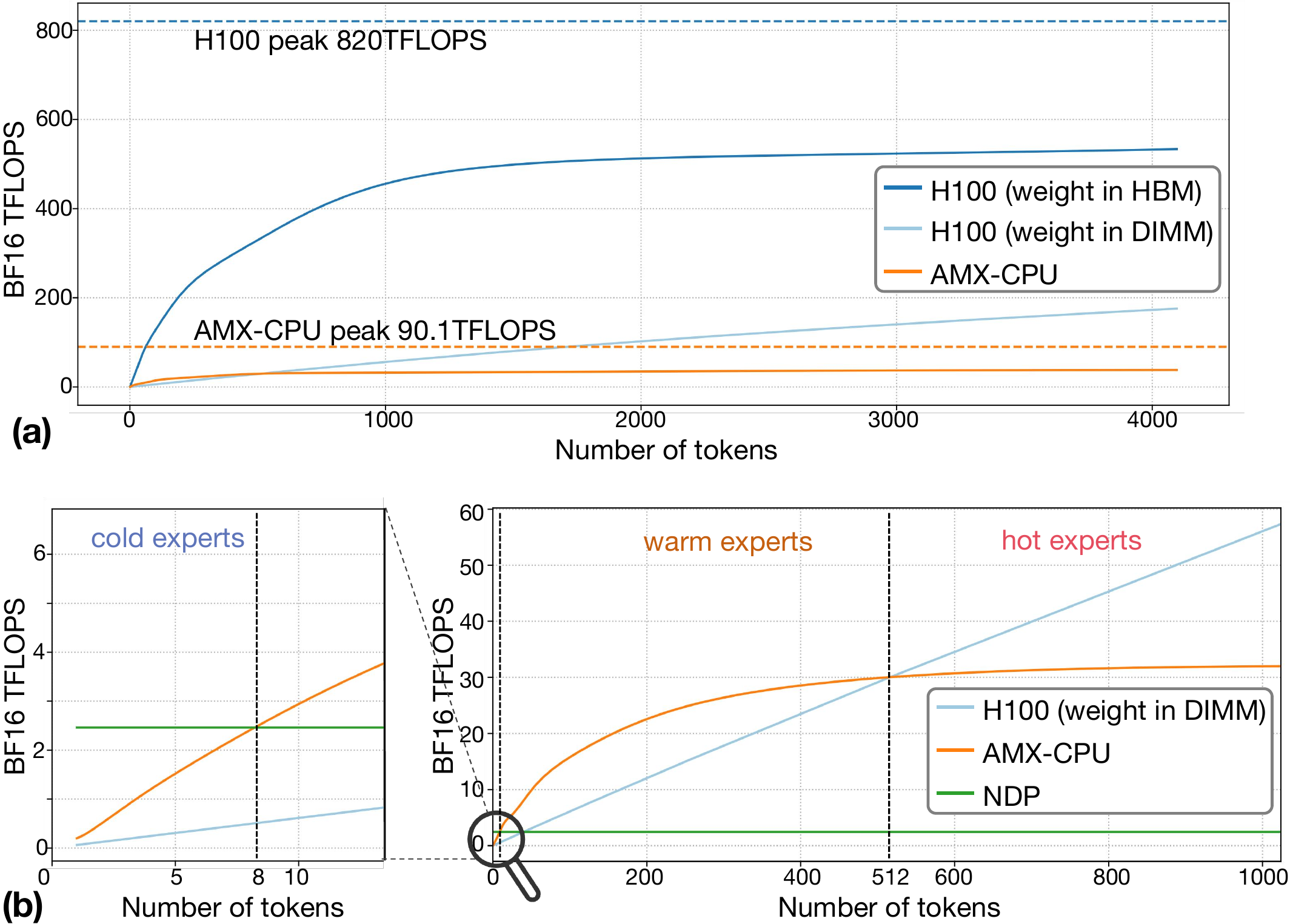}
\caption{Compute Characterization. (a) Measured Throughput vs. Token Count. (b) Empirical GPU-CPU-NDP roofline.}
\label{fig:roofline}
\end{figure}

SOTA GPU-NDP systems assume that all non-hot experts form a uniform, memory-bandwidth-bound subset suitable for DIMM-NDP processing~\cite{kim2024monde, wu2025pimoe}. However, our analysis shows this is an oversimplification. As illustrated in Figure~\ref{fig:hot-warm-cold}, an activation study across multiple representative MoE models and real-world datasets reveals substantial heterogeneity among non-hot experts: a long tail of cold experts constitutes over 70\% of all experts but processes only 8\% of tokens, while 20\%–40\% of warm experts handle up to 70\% of tokens. This heterogeneity creates a scheduling dilemma for GPU-NDP: (1) \textbf{GPU-side I/O bottleneck}. This bottleneck is two-fold. First, as shown in Figure~\ref{fig:roofline}(a), the H100 GPU requires at least 256 tokens per expert to reach 30\% utilization, even when the expert is resident in HBM. Warm experts fall well below this threshold. More critically, these non-resident warm experts must be fetched via PCIe. Their low token count yields computation times too brief to overlap the significant I/O latency, forcing the GPU to stall. Consequently, overall GPU utilization drops below 8.6\%. (2) \textbf{NDP-side compute bottleneck}. Scheduling warm experts on the NDP exceeds its compute budget, increasing NDP execution time by 7$\times$ and offsetting the benefits of cold-expert offloading.

\subsection{Opportunity: AMX-CPU Bridging the GPU–NDP Compute Gap}
The scheduling dilemma in Section~\ref{Sec:Dilemma-of-warm} arises from a significant compute gap between the high throughput of GPU and the limited compute of NDP. Warm experts fall squarely into this gap, which prompts a re-examination of CPU capability that NDP-based designs have largely overlooked. Although Section~\ref{Sec:MoE-offloading} notes that GPU-CPU systems are constrained by memory bandwidth, we argue that the root cause is the misallocation of CPU to bandwidth-intensive cold experts for which CPU is ill-suited. Server-class CPUs with matrix extensions, such as Intel AMX and ARM SME, can deliver 10–40 TFLOPS when executing warm experts (processing tens to hundreds of tokens)~\cite{chen2025ktransformers}, effectively outperforming both the PCIe-bound GPU and the compute-limited NDP as shown in Figure~\ref{fig:roofline}(b). This observation leads to a clear three-way co-design: AMX-enabled CPU handles warm experts to fill the compute gap; DIMM-NDP units process long-tail cold experts by exploiting their high internal bandwidth to avoid the CPU’s memory bottleneck; and the GPU continues to serve hot experts efficiently. This division assigns each resource to the task it executes most effectively.

\textbf{Challenges:} 
However, realizing this co-design poses two central challenges: \textit{(1) Orchestration Complexity:} How to design a lightweight scheduler that dynamically adapts to fluctuating expert loads and optimizes placement across three heterogeneous domains with distinct compute-I/O trade-offs. \textit{(2) Data Management Conflict:} How to design an adaptive data placement policy that reconciles the structural conflict between layout preferences~\cite{liu2025l3, zhao2024pim} (striped for CPU/GPU vs. localized for NDP), while simultaneously mitigating the NDP load skew caused by uneven cold expert activations~\cite{kim2025lia}.

\section{TriMoE}


This section details the design and implementation of TriMoE. We first present the overall architecture (Section~\ref{Sec:TriMoE-arch}), followed by two novel policies that address the challenges: the Bottleneck-Aware Greedy Makespan Expert Scheduling (Section~\ref{Sec:scheduler}) and the Prediction-Driven Expert Relayout and Rebalancing (Section~\ref{Sec:relayout}).

\subsection{TriMoE Architecture}\label{Sec:TriMoE-arch}

Figure~\ref{fig:system} illustrates the overall TriMoE architecture, which augments a server-grade single-GPU system with three heterogeneous compute resources: a high-performance GPU, an AMX-enabled CPU, and DIMM-NDP, to achieve high-throughput MoE offloading.

\textbf{High-Performance GPU:} With hundreds of TFLOPS~\cite{choquette2021nvidia, choquette2023nvidia}, the GPU serves as the core compute unit for high-density tasks. During the prefill phase, it performs all inference computations. In the decode phase, it focuses on MLP, Attention, and the computation of hot experts prefetched into HBM. For storage, MLP and shared expert weights reside in GPU HBM, while the large KV Cache and all routed experts are offloaded to host DIMMs~\cite{xu2025moe}.

\textbf{AMX-enabled CPU:} CPUs with matrix extensions provide considerable compute power. For example, an Intel Xeon Sapphire Rapids (90.1 TFLOPS theoretical) achieves 22\% of a measured A100 GEMM throughput~\cite{kim2025lia}. Unlike the GPU which relies on remote PCIe transfers, the CPU operates directly on host-resident data. TriMoE exploits this property by designating the AMX-CPU as a dedicated compute domain for warm experts, maximizing the utility of host-side resources that would otherwise be bottlenecked by memory bandwidth if assigned to cold experts.


\textbf{DIMM-NDP:} To process long-tail cold experts, we adopted a center buffer-based NDP instead of a high-area-overhead bank-level NDP~\cite{alian2018application, ke2020recnmp, cong2017aim, park2024attacc}. As shown in Figure~\ref{fig:system}(a), the NDP units are placed on the DIMM's Buffer Chip. This design is an efficient trade-off: it leverages high aggregate internal bandwidth (8$\times$ host bandwidth) to efficiently process cold experts and remains fully compatible with regular memory access, without interfering with normal CPU or GPU access to DIMM data~\cite{alian2018application, cong2017aim}.

\textit{GEMV \& Act Unit:} Each DIMM-NDP is equipped with a dedicated GEMV unit. It contains 256 multipliers and a multi-level adder tree. Each 128-bit multiplier operates in a typical bit-serial manner~\cite{devaux2019true}, processing 8 FP16 values simultaneously, while a 256KB internal buffer is used to store intermediate activations. To support non-linear functions such as SiLU in MoE layers, the unit is further integrated with an activation module that primarily comprises 256 FP16 exponentiation and multiplication units.

\textit{Relayout Unit \& DIMM-Link:} A core architectural challenge is data layout preference conflict and the NDP load imbalance. To resolve both challenges, we need the ability to perform fast data relayout and migration between DIMMs, as detailed in Section~\ref{Sec:relayout}. We therefore adopt DIMM-Link~\cite{zhou2023dimm} and introduce a Relayout Unit. The DIMM-Link provides a 25 GB/s host-free, cross-DIMM transfer bus. The Relayout Unit acts as a control engine, leveraging the DIMM-Link to efficiently execute two key background tasks. (1) Relayout: enabling data conversion between striped and localized layouts, and (2) Rebalancing: enabling fast migration of localized cold experts between DIMMs. This host-free data operation is over 62$\times$ more efficient than CPU-driven data movement.

\textbf{Scheduler:} A scheduler orchestrates all the heterogeneous resources, performing online expert scheduling and dynamic expert relayout and rebalancing, detailed in subsequent sections.

\subsection{Bottleneck-Aware Greedy Makespan Expert Scheduling}\label{Sec:scheduler}

Scheduling on TriMoE's three-way heterogeneous platform is challenging. While mapping experts to their best-suited device aims to maximize resource utilization, this simple theoretically optimal strategy can create global bottlenecks. The key is to balance the load across all three domains to minimize the global makespan. We propose a two-phase scheduling policy: it first performs a greedy, cost-model-based initial assignment, followed by an iterative, bottleneck-aware refinement to optimize the final makespan.

\textbf{Expert Execution Cost Model:} The first step is to estimate the execution cost of any expert $E_i$ on all possible paths and select the most suitable one. This cost model relies on two runtime inputs, as shown in Figure~\ref{fig:system}(b): the expert load (token count) $L_i$ from the gate function and the expert's mapping and data layout, collectively represented by $M_i$. To accurately model compute latency, we offline-profile both the GPU and CPU for various token counts, building lookup tables~\cite{ke2022hercules} for $f_{calc\_gpu}(L_i)$ and $f_{calc\_cpu}(L_i)$.

\textit{1) GPU Execution Cost:} We evaluate two paths. Equation~\ref{eq:gpu_hit} models a cache hit, where the expert is resident in GPU HBM and the cost is dominated by computation. Equation~\ref{eq:gpu_miss} represents a cache miss, requiring on-demand loading from the host. its cost is determined by GPU computation, PCIe weight transfer, and host DIMM weight read ($T_{DRAM}$). The $T_{DRAM}$ time depends on the expert's weight layout ($M_i$), which is either striped, accessing total memory bandwidth, or localized, accessing only single-DIMM bandwidth.
\begin{align}
T_{GPU\_Hit}(E_i) &= f_{calc\_gpu}(L_i) \label{eq:gpu_hit} \\
T_{GPU\_Miss}(E_i) &= \max\left( f_{calc\_gpu}(L_i), T_{PCIe}, T_{DRAM}(W_i, M_i) \right) \label{eq:gpu_miss}
\end{align}

\textit{2) CPU Execution Cost:} For the AMX-CPU, the cost is the maximum of its computation time and the DIMM weight read time ($T_{DRAM}$), as $T_{DRAM}$ is similarly dependent on the expert's layout.
\begin{equation}
T_{CPU}(E_i) = \max\left( f_{calc\_cpu}(L_i), T_{DRAM}(W_i, M_i) \right)
\label{eq:cpu_cost}
\end{equation}


\textit{3) NDP Execution Cost:} Unlike the centralized GPU and CPU, DIMM-NDP units operate in parallel. We restrict this path strictly to experts with a localized layout to avoid high overhead from input broadcasting and result reduction across multiple DIMMs. Consequently, an expert $E_i$ is only executable on the specific DIMM where its weights reside. Its cost is the maximum of NDP computation and the time to read weights using the internal DIMM bandwidth.
\begin{equation}
T_{NDP}(E_i) = \max\left( f_{calc\_ndp}(L_i), T_{Internal}(W_i) \right)
\label{eq:ndp_cost}
\end{equation}


\textbf{Overall Makespan Modeling:} After the initial expert assignment, we formulate the global makespan by aggregating costs within each compute domain. Let $\mathcal{S}_{GPU}$, $\mathcal{S}_{CPU}$, and $\mathcal{S}_{NDP}^d$ denote the sets of experts assigned to the GPU, CPU, and the $d$-th DIMM-NDP unit, respectively. 

The GPU and CPU act as centralized serial processors. Their total execution times, $T_{total}^{GPU}$ and $T_{total}^{CPU}$, are simply the cumulative sum of the execution costs of their assigned experts (derived from Eq.~\ref{eq:gpu_hit}--\ref{eq:cpu_cost}):
\begin{equation}
T_{total}^{GPU/CPU} = \sum_{E_i \in \mathcal{S}_{GPU/CPU}} T_{GPU/CPU}(E_i)
\label{eq:domain_serial}
\end{equation}
In contrast, the DIMM-NDP units operate in parallel, so the NDP domain latency is dictated by the bottleneck DIMM. Crucially, our model accounts for \textit{memory access contention}: a DIMM is busy not only when performing local NDP computation but also when serving weight fetch requests from the GPU or CPU. The total active time for a DIMM $d$ is:
\begin{equation}
T_{DIMM}(d) = \sum_{E_i \in \mathcal{S}_{NDP}^d} T_{NDP}(E_i) + T_{contention}^d
\label{eq:dimm_load}
\end{equation}
where $T_{contention}^d$ represents the accumulated DRAM access latency caused by striped or localized weight reads from the host processors targeting DIMM $d$. The global makespan is ultimately determined by the slowest of the three domains:
\begin{equation}
T_{Makespan} = \max \left( T_{total}^{GPU}, T_{total}^{CPU}, \max_{d \in \mathbb{D}} T_{DIMM}(d) \right)
\label{eq:global_makespan}
\end{equation}

\textbf{Bottleneck-Aware Refinement:} The initial expert assignment may create a global load imbalance, so the algorithm enters an iterative bottleneck correction phase. In each iteration, the algorithm first identifies the bottleneck device (the one with the maximum total time) and selects the highest-cost expert on that bottleneck device as a migration candidate. It then evaluates re-assigning this expert to the other two devices and models the new global makespan for each potential move. The algorithm greedily selects the move that results in the minimum new global makespan. If both moves result in the same new makespan, the algorithm breaks the tie by choosing the move that causes the minimum time increase (delta) on the receiving device. This iterative refinement continues until no re-assignment can further reduce the global makespan, or a maximum iteration limit is reached.

\subsection{Prediction-Driven Expert Relayout and Rebalancing}\label{Sec:relayout}

The effectiveness of the online expert scheduling is constrained by expert weight placement. Although we set an optimized initial layout via offline trace analysis (e.g., localizing cold experts onto single DIMMs), this static placement becomes suboptimal at runtime due to dynamic expert activation~\cite{he2025hydra, yu2025fmoe}. This leads to two core problems. First, a \textbf{Layout Preference Conflict} arises as an expert's identity can change. For instance, a cold expert may become warm, creating a mismatch with the CPU's preference for a striped layout. Second, \textbf{NDP Load Imbalance} occurs when uneven activation of cold experts across DIMMs overloads certain DIMM-NDP, creating bottlenecks. To address these issues, we design a prediction-driven policy that leverages the temporal locality of expert activation to perform asynchronous relayout and rebalancing~\cite{xue2024moe, fang2025accurate}.

\textbf{Expert Load Predictor:} As shown in Figure~\ref{fig:system}(b), we introduce a lightweight Expert Load Predictor. The predictor maintains an Exponential Moving Average (EMA) for each expert's load~\cite{zeng2025efficientmoe,haynes2012exponential}, which is updated after every decode step:
\begin{equation}
EMA_e(t) = \alpha \cdot F_e(t) + (1-\alpha) \cdot EMA_e(t-1)
\label{eq:ema}
\end{equation}
where $F_e(t)$ is the expert's actual load at step $t$ and $\alpha=0.3$ is empirically tuned to predict the expert's load trend while suppressing noise, yielding >78\% accuracy.


\textbf{Expert Migration Decision:} When one layer's expert computation completes, the predictor estimates the next layer's load trends and triggers three types of background actions. \textit{(1) Hot Expert Prefetching:} If an expert is predicted as hot, the runtime triggers a high-priority task to prefetch its weights via PCIe to GPU HBM. \textit{(2) Dynamic Relayout:} If an expert's layout mismatches its optimal execution unit, the Relayout Unit uses the DIMM-Link to convert its data layout between localized and striped. \textit{(3) Cold Expert Rebalancing:} To resolve NDP load imbalance, the system calculates the total predicted load of cold experts on each DIMM. If skew is detected, it greedily selects and migrates cold experts one by one from the busiest DIMM to the most idle DIMM via DIMM-Link. The overheads of these migrations are intended to be hidden within the overlap window provided by the current layer's MLP and Attention computations on the GPU. Our policy ranks all feasible migration tasks by their predicted benefit and greedily executes them in priority order until their cumulative estimated time fills this window budget. This ensures the limited DIMM-Link bandwidth is spent on the most critical tasks. 

Our evaluation shows that the predictor achieves 78\% migration decision accuracy with only 38 KB of memory overhead to store necessary metadata, yielding a 1.16$\times$ end-to-end performance gain.
\section{Evaluation}

\subsection{Experiment Setup}

\subsubsection{TriMoE System} To evaluate TriMoE, we built a heterogeneous prototype system comprising an NVIDIA H100 PCIe GPU with 80GB HBM, an AMX-enabled Intel Xeon Platinum 8470 CPU (8-channel memory), and 16 DIMM-NDPs providing high internal bandwidth. Detailed configurations are listed in Table~\ref{table:system_config}. We utilize PCIe 5.0 to provide 64GB/s unidirectional bandwidth for host-to-GPU data transfer. On the software stack, we employ vLLM 0.8.1~\cite{kwon2023efficient} and KTransformers~\cite{chen2025ktransformers} to implement efficient kernels for the GPU and AMX-CPU, respectively. We developed a cycle-accurate simulator based on Ramulator 2.0~\cite{luo2023ramulator} to evaluate DIMM-NDP performance. For NDP units, we implemented them using RTL and synthesized it using the Synopsys Design Compiler~\cite{sdc} with the TSMC 7nm technology.


\begin{table}[htbp]
\centering
\caption{System configurations.}
\renewcommand{\arraystretch}{1} 
\scriptsize
\begin{tabular}{@{} >{\centering\arraybackslash}m{1.1cm} >{\centering\arraybackslash}m{6.9cm} @{}}
\toprule
\multirow{1}{*}{\textbf{GPU}} 
    & H100 80GB: 819.6 TFLOPS (BF16), 2.04 TB/s Bandwidth, 64 GB/s PCIe bandwidth \\
\midrule
\multirow{2}{*}{\textbf{CPU}} 
    & Memory: 8 Channels $\times$ 2 DIMM, 2TB Capacity \\
    & AMX: 90.1 TFLOPS (BF16), 307.2 GB/s Bandwidth \\
\midrule
\multirow{7}{*}{\textbf{DIMM-NDP}} 
    & \textbf{DIMM Parameters} \\
    & DDR5-4800, 128GB/DIMM, 4 Ranks $\times$ 8 Bank Group $\times$ 4 Banks \\
    & CCDL=12:CCDS=8:RRDS=8:RRDL=12:FAW=32:RP=34:BL=8:CL=40:WR=72\\
\cmidrule(l){2-2}
    & \textbf{NDP} \\
    & 256 GFLOPS (BF16) \& 153.6 GB/s Bandwidth, $1.13mm^2$ area overhead per NDP \\
\cmidrule(l){2-2}
    & \textbf{DIMM-Link Parameters} \\
    & 25Gb/s/Lane, 8 $\times$ Lanes (25GB/s per Link) \\
\bottomrule
\end{tabular}
\label{table:system_config}
\end{table}

\subsubsection{Baselines} We compare TriMoE with three SOTA MoE offloading systems. (1) Klotski~\cite{fang2025klotski}: GPU-Only. A baseline that maximizes PCIe overlap by prefetching hot experts and loading cold expert weights during hot expert computation; (2) Enhanced Ktransformer~\cite{chen2025ktransformers}: GPU-CPU. We added prefetching and on-demand loading of hot experts on top of its processing of all non-shared experts using AMX-CPU, making it the strongest baseline under this architecture~\cite{zhong2025hybrimoe}; (3) MoNDE~\cite{kim2024monde}: GPU-NDP. A system that balances load between GPU and NDP by modeling the cost trade-off between weight migration (to GPU) and activation migration (to NDP). 

\subsubsection{Workloads \& Metrics} We select three representative modern MoE models with specifications detailed in Table~\ref{table:model_spec} using LMSys~\cite{zheng2023lmsys} and CodeAlpaca~\cite{huggingfaceh4_codealpaca_20k} datasets to extract real expert activation traces to construct workloads. As offloading systems primarily target high-throughput demands~\cite{xu2025moe}, we focus on large batch sizes of 256–768 aggregated via zigzag or offline batching, supplemented by robustness assessments on small batches. Given that the MoE decode phase dominates over 90\% of inference time, we report MoE layer latency and end-to-end throughput as our core metrics.


\begin{table}[htbp]
\centering
\caption{MoE models configurations.}
  \scriptsize
  \setlength{\linewidth}{0pt}   
  \renewcommand{\arraystretch}{1}
  \begin{tabularx}{\columnwidth}{@{}%
      l!{\vrule\vrule}%
      c!{\vrule\vrule}%
      c|c|c|c%
      @{}}
    \Xhline{1.1pt}
    \textbf{Model} & \textbf{params} & \textbf{shared exp} & \textbf{routed exp} & \textbf{top-$k$} & \textbf{expert params} \\
    \Xhline{0.8pt}
    DeepSeek-V2~\cite{liu2024deepseek}  & 236B & 2 & 160 & 6 & 422GB \\
    Qwen3-235B-A22B~\cite{yang2025qwen3} & 235B & 0 & 128 & 8 & 423GB \\
    GLM-4.5-Air~\cite{zeng2025glm}  & 106B & 1 & 128 & 8 & 190GB \\
    \Xhline{1.1pt}
  \end{tabularx}
   \label{table:model_spec}
\end{table}

\subsection{TriMoE Performance}

\subsubsection{MoE Layer Decode Speedup} Our evaluation targets the MoE decode phase, the primary bottleneck in inference. As shown in Figure 6, TriMoE achieves an average 2.12–2.83$\times$ speedup over the best baseline across all models. To identify the source of these gains, we analyze resource utilization (Table~\ref{table:resource_util}). Klotski is constrained by severe PCIe bottlenecks, resulting in only 28.6\% GPU utilization. En-KTransformers is limited by host bandwidth when serving cold experts on the CPU, capping CPU utilization at 42\%. Meanwhile, MoNDE’s limited NDP compute capability forces the GPU to retain a large fraction of expert computation, failing to relieve PCIe congestion. In contrast, TriMoE offloads warm experts to AMX-enabled CPU and assigns cold experts to NDP, sustaining an average utilization of 76.2\% across the three compute domains.

\begin{table}[htbp]
  \centering
  \caption{Resource utilization comparison}
  \scriptsize
  \begin{tabular}{l|cccc} 
    \Xhline{1.1pt}
    Resource & Klotski & En-KTransformers & MoNDE & \textbf{ours} \\
    \Xhline{0.8pt}
    GPU Compute Utilization & 28.6\% & 57.6\% & 33.9\% & \textbf{66.0\%} \\
    CPU Compute Utilization & $\times$ & 42\% & $\times$ & \textbf{74.9\%} \\
    NDP Compute Utilization & $\times$ & $\times$ & 70.1\% & \textbf{87.8\%} \\
    \Xhline{1.1pt}
  \end{tabular}
  \label{table:resource_util}
\end{table}

\begin{figure}
\centering
\includegraphics [width=1.0\linewidth]{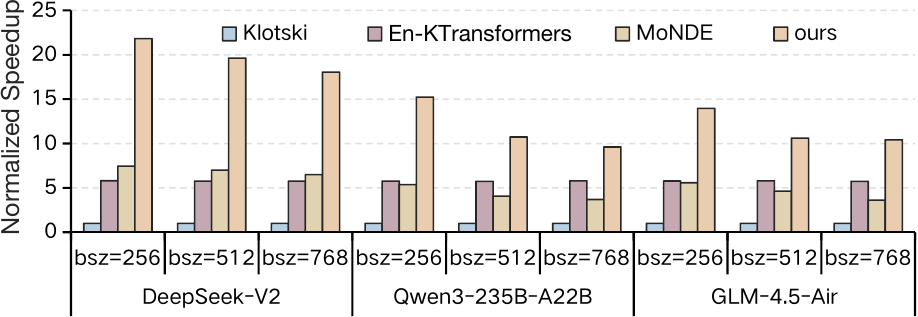}
\caption{MoE decode speedup over Klotski.}
\label{fig:speedup}
\end{figure}

\subsubsection{End-to-End Performance}

\begin{figure}
\centering
\includegraphics [width=1.0\linewidth]{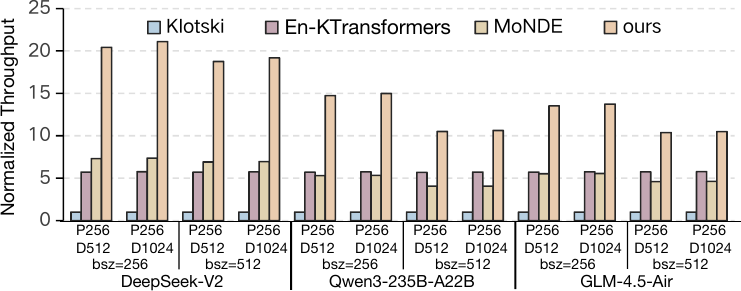}
\caption{End-to-End throughput over Klotski.}
\label{fig:throughput}
\end{figure}

We further evaluate end-to-end performance. As shown in Figure~\ref{fig:throughput}, TriMoE achieves 2.78$\times$, 2.22$\times$, and 2.09$\times$ higher throughput over the strongest baseline across the three models. This confirms that the gains at the MoE decode layer translate effectively into system-level performance improvements, becoming notably more pronounced as output length increases.

\subsection{Ablation Study}

\begin{figure}
\centering
\includegraphics [width=.98\linewidth]{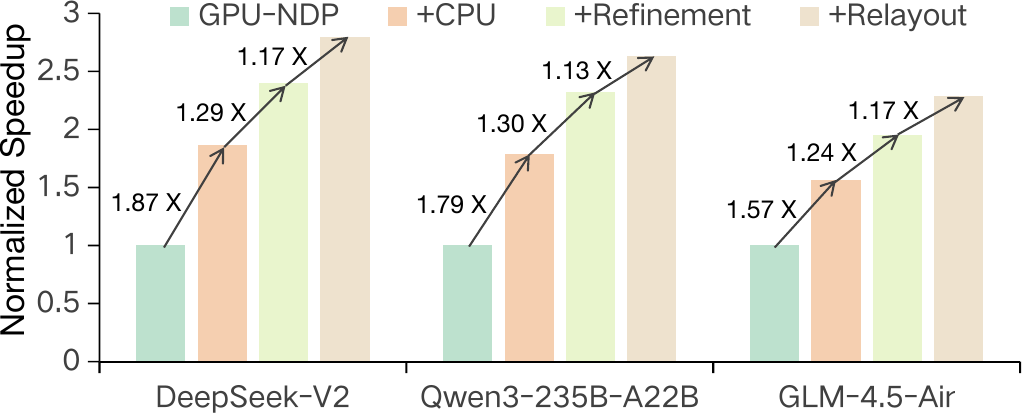}
\caption{Ablation study of TriMoE with batch size 512.}
\label{fig:ablation}
\end{figure}

As shown in Figure~\ref{fig:ablation}, we quantify component contributions via an ablation study starting from a \textbf{GPU-NDP} baseline. First, incorporating AMX-CPU with greedy scheduling (\textbf{+CPU}) yields the most significant 1.75$\times$ speedup. This validates our hypothesis that AMX-CPU effectively processes warm experts, breaking the deadlock between GPU I/O and NDP compute bottlenecks. Next, enabling Bottleneck-Aware Refinement (\textbf{+Refinement}) contributes a further 1.28$\times$ gain, confirming that the algorithm corrects greedy mapping imbalances to minimize the global makespan. Finally, Prediction-Driven Relayout and Rebalancing (\textbf{+Relayout}) adds another 1.16$\times$ improvement. With a predictor accuracy exceeding 78\%, this mechanism successfully resolves runtime layout conflicts and NDP load skew caused by dynamic activations, ensuring sustained system efficiency.

\subsection{Sensitivity Analysis}

\subsubsection{Number of NDPs} Figure~\ref{fig:sensitivity}(a) varies the number of NDP-equipped DIMMs. Latency stabilizes after 16 units. This reflects a key workload characteristic: cold experts, though numerous, possess low aggregate compute density. Thus, a full-NDP configuration provides sufficient bandwidth to process all cold experts.

\subsubsection{CPU Compute Capability} 
\begin{figure}
\centering
\includegraphics [width=1.0\linewidth]{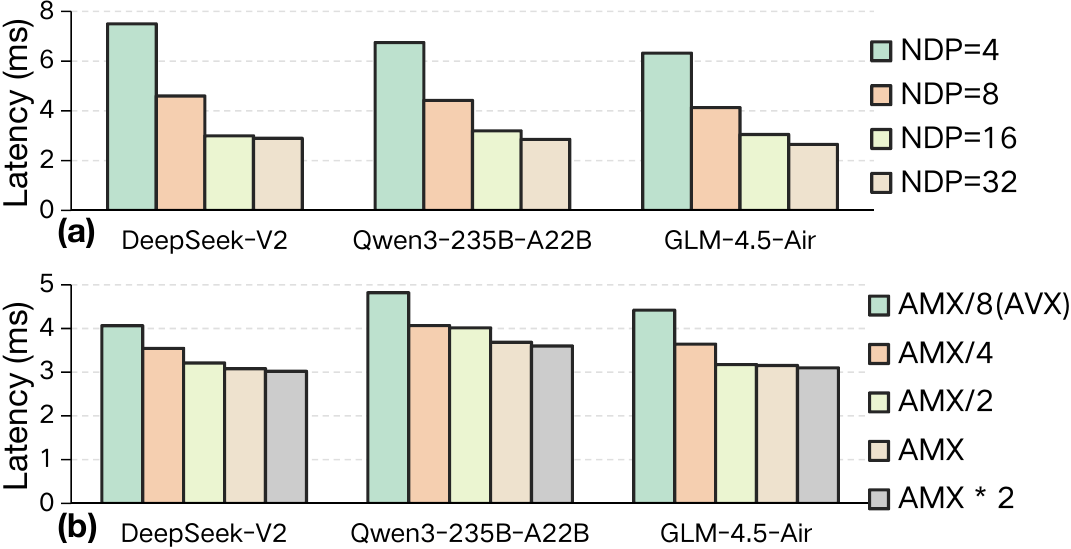}
\caption{Sensitivity study of NDP count and CPU TFLOPS.}
\label{fig:sensitivity}
\end{figure}

Figure~\ref{fig:sensitivity}(b) sweeps CPU compute power from legacy AVX levels ($\approx$0.125$\times$) to 2.0$\times$ AMX. While AVX levels exhibit higher latency, the curve flattens significantly once capacity reaches 0.5$\times$ of the AMX baseline. This confirms that modern AMX extensions offer sufficient computational capacity, allowing even mid-range CPUs to effectively bridge the warm expert gap.

\subsection{Robustness \& Overhead}

To evaluate robustness using the Qwen model at batch sizes 128, 64, and 32, TriMoE maintains speedups of 2.72$\times$, 2.18$\times$, and 1.82$\times$ over the strongest baseline, confirming its efficacy across varying throughput demands. We also analyzed the runtime overhead. Leveraging the DIMM-Link's host-free, parallel inter-dimm transfer capability, we find that the $\sim$0.63 ms cost of migrating up to four experts is entirely masked by the $\sim$0.68 ms concurrent GPU computation. Given the predictor accuracy exceeding 78\% (which makes such large transfers infrequent), the overall online migration overhead measured remains limited to below 3.3\%.

\section{Conclusion}

TriMoE revisits MoE offloading from the perspective of expert heterogeneity and exposes the fundamental inefficiency of binary GPU–NDP designs in handling warm experts. By jointly leveraging GPU for hot experts, AMX-enabled CPU for warm experts, and DIMM-NDP for cold experts, TriMoE matches each expert class to its most suitable compute domain. Combined with our bottleneck-aware scheduling and prediction-driven relayout and rebalancing, this tri-domain architecture achieves up to 2.83$\times$ speedup over state-of-the-art offloading systems.
\bibliographystyle{ACM-Reference-Format}
\bibliography{references}

\end{document}